\documentclass[twocolumn,floatfix,showpacs,showkeys,superscriptaddress]{revtex4-1}
\usepackage{graphicx}
\usepackage{amssymb}
\usepackage{amsmath}
\usepackage{bm}
\begin{document}
\setcounter{page}{0}
\title[]{Traveling Baseball Players' Problem in Korea}
\author{Hyang Min \surname{Jeong}}
\affiliation{Department of Physics and BK21 Physics Research Division,
Sungkyunkwan University, Suwon 440-746, Korea}
\author{Sang-Woo \surname{Kim}}
\affiliation{Department of Physics, Soongsil University, Seoul 156-743, Korea}
%\thanks{Present address: Department of Physics, Soongsil University, Seoul 156-743, Korea}
%\affiliation{Department of Physics, University of Seoul, Seoul 130-743, Korea}
\author{Aaram J. \surname{Kim}}
\affiliation{Department of Physics and Astronomy and Center for Theoretical Physics, Seoul National University, Seoul 151-747, Korea}
\author{Younguk \surname{Choi}}
\affiliation{Department of Physics, Soongsil University, Seoul 156-743, Korea}
\author{Jonghyoun \surname{Eun}}
\affiliation{Department of Physics and Astronomy, University of California Los Angeles, Los Angeles, California 90095-1547, USA}
\author{Beom Jun \surname{Kim}}
\email[Corresponding author: ]{beomjun@skku.edu}
\affiliation{Department of Physics and BK21 Physics Research Division,
Sungkyunkwan University, Suwon 440-746, Korea}

\date{\today}

\begin{abstract}
We study the so-called the traveling tournament problem (TTP),  to find an optimal tournament schedule. Differently
from the original TTP, in which the total travel distance of  all the
participants is the objective function to minimize, we instead seek to maximize
the fairness of the round robin tournament schedule of the Korean Baseball
League. The standard deviation of the travel distances of teams is defined
as the energy function, and the Metropolis Monte-Carlo method combined
with the simulated annealing technique is applied to find the ground
state configuration. The resulting tournament schedule is found to satisfy
all the constraint rules set by the Korean Baseball Organization, but
with drastically increased fairness in traveling distances.

\end{abstract}

%89.65.-s	Social and economic systems
%89.75.Fb	Structures and organization in complex systems


\maketitle

\section{Introduction}
\label{sec:intro}

The traveling salesman problem (TSP) is the one of the most
important optimization problems in computational sciences~\cite{TSPbook}. 
In TSP, for given number of places to visit, a traveling salesman hopes
to minimize the total travel distance within the constraint that 
she is allowed to visit each place only once.
Hinted by the similarity to the problem of finding the ground state of  a
system with a rugged energy landscape, TSP has also been broadly studied in physics
community~\cite{kirkpatrick,leechoi}. There are also biologically motivated
interesting methods in the study of TSP, such as the genetic algorithm~\cite{bradynature} and
the ant colony optimization~\cite{ACObook,dorigo}.
Although direct formulations based on TSP have broad range of applicability, we note that
there are closely-related, but different, problems in reality. One among such problems is the scheduling of games in 
sports: In contrast to TSP, which is unilateral, i.e., the traveling salesman 
is not required to take into account existences of other salesmen, many sports games are bilateral, i.e.,
two teams or players should meet somewhere to play a game. 
Designing of a tournament schedule in sports league is notoriously
difficult problem to solve, which has been studied mostly in the area of computational 
science~\cite{schedulebook,kendall}.  
Recently, there has been a significant development in the traveling tournament
problem (TTP)~\cite{eastonTTP}, in which the task is to construct the optimal
timetable to minimize the total travel distance under the constraint of a given
home-and-away pattern. Not surprisingly, the simulated-annealing method~\cite{TTP:SA},
and the ant colony optimization method~\cite{TTP:ACO} have been applied,
as well as more conventional methods in computer science~\cite{trick,*schaerf},
to tackle TTP and similar problems.

\begin{table}
\caption{Korean Baseball League of eight teams. Note that
two teams (Doosan and LG) share the same home stadium (Jamsil).
The latitude and the longitude are for the locations
of stadiums. Acronyms used in the present work are also listed.}
\begin{tabular}{c|c|c|c}
\toprule
\# & team name & stadium & (latitude, longitude) \\
\colrule
 1 &Samsung Lions (SS) &  Daegu    & (35.882, 128.585) \\
 2 &SK Wyverns    (SK) &  Munhak      & (37.436, 126.689) \\
 3 &Lotte Giants  (LT) &  Sajik     & (35.195, 129.059) \\
 4 &Kia Tigers    (KA) &  Gwangju     & (35.168, 126.888) \\
 5 &Doosan Bears  (DS) &  Jamsil    & (37.514, 127.075) \\
 6 &LG Twins      (LG) &  Jamsil        & (37.514, 127.075) \\
 7 &Hanwha Eagles (HH) &  Daejeon & (36.314, 127.428) \\
 8 &Nexen Heroes  (NX) &  Mokdong  & (37.531, 126.881) \\
\botrule
\end{tabular}
\label{tab:teams}
\end{table}

\begin{figure}
\includegraphics[width=0.45\textwidth]{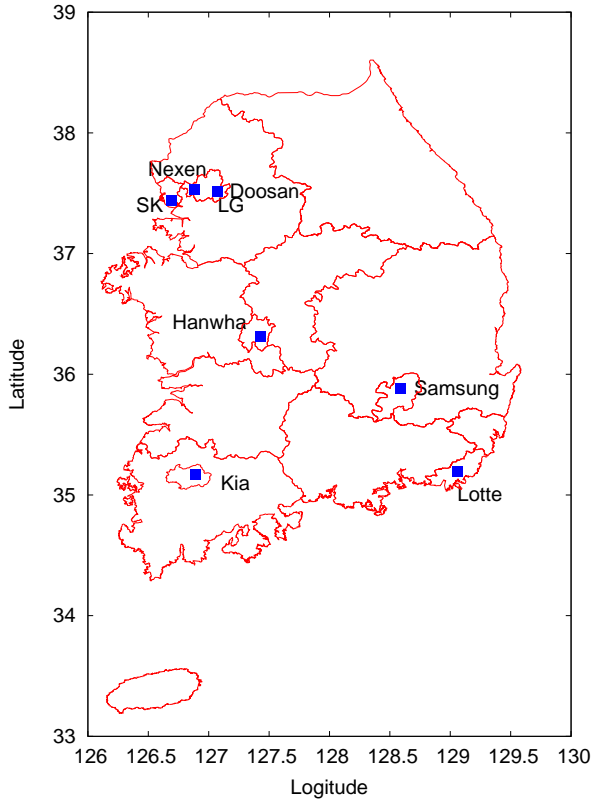}
\caption{(Color online) Locations of 8 baseball teams in Korean Baseball League. Two teams,
Doosan Bears and LG Twins, share the same home stadium. Lotte is located 
at the farthest distance from Seoul.} 
\label{fig:stadium}
\end{figure}

In this work, we study the TTP, but with an important alteration: Instead of
minimizing the total travel distance, we minimize {\em the unfairness of the
travel distance}. This, we believe, is even more important than just minimizing
the total travel distance summed over all the teams in the sports league. 
If one team should travel much longer distance than others, the team has
less chance to win the tournament. Consequently, for the sake of fairness, 
the difference of the travel distance of each team must be made as small as 
possible, even though it might sacrifice the total sum of the travel distances of all the teams.
In this respect, the present work is related with discussions of social
justice in which equality of opportunity is the one of the key 
issues~\cite{rawls,*dworkin,*skbaek}.
Specifically, this work is motivated by the
common belief among baseball fans in Korea. One team (Lotte Giants) is located
in the most southern part of Korean peninsula, and thus it is not fair for
the team to compete with all others on the same ground.  Interestingly, this
claim,  mostly by fans of Lotte Giants -- they are also well known to be very
enthusiastic by the way -- has some truth in it, and the Korean Baseball
Organization (KBO) set up a rule
trying to reduce the travel distance of the Lotte Giants: KBO allows the team
to have more away games in a row without returning to their home (see Sec.~\ref{sec:kborule}).

The present paper is organized as follows: In Sec.~\ref{sec:kborule}, we describe
in detail the current scheduling scheme of KBO, which is followed by 
the formulation of the problem in terms of the Monte-Carlo method in statistical
physics in Sec.~\ref{sec:form}. The results are presented in Sec.~\ref{sec:res},
which are summarized in Sec.~\ref{sec:sum} with some discussion and proposal.

\begin{table*}
\caption{Distance matrix for baseball teams. The number at the $i$th row and $j$th column
denotes the distance $d_{ij}$ from $i$ to $j$ in units of km, and the average distances ${\bar d_{i \bullet}}$
and ${\bar d_{\bullet j }}$ are defined by ${\bar d_{i\bullet}} \equiv (1/8)\sum_{j=1}^8 d_{ij}$ and 
${\bar d_{\bullet j}} \equiv (1/8)\sum_{i=1}^8 d_{ij}$.  For the team index $i$, see
Table~\ref{tab:teams}. }
\begin{tabular}{c|c|c|c|c|c|c|c|c}
\toprule
from\textbackslash to & 1 (SS)  &  2 (SK)  &  3 (LT)  &  4 (KA)  &  5,6 (DS,LG)  &  7 (HH)  &  8 (NX)   & ${\bar d_{i \bullet}}$ \\  
\colrule
 1 (SS) & 0      & 299.0 & 106.7 & 220.0 & 279.2 & 147.4 & 298.2 & 203.7 \\ \colrule                                    
 2 (SK) & 298.6 & 0      & 404.8 & 315.3 & 54.6  & 178.5 & 30.9 & 167.1 \\ \colrule                                    
 3 (LT) & 105.7 & 404.9 & 0      & 254.9 & 379.6 & 253.1 & 404.1 & 272.7 \\ \colrule                                    
 4 (KA) & 215.6 & 309.5 & 253.9 & 0      & 294.2 & 174.7 & 308.6 & 231.3\\ \colrule                                    
 5,6 (DS,LG) & 278.8 & 53.1  & 385.1 & 290.5  & 0     & 162.6 & 23.2  & 149.16 \\ \colrule                                    
 7 (HH) & 147.3 & 181.9 & 253.6 & 175.4  & 174.9 & 0      & 181.2 & 159.2\\ \colrule                                    
 8 (NX) & 301.9 & 29.4  & 408.1 & 312.7  & 22.3   & 184.8 & 0   & 180.0  \\ \colrule
 ${\bar d_{\bullet j}}$ & 203.4 & 166.4 & 274.7 & 232.4 & 150.6 & 158.0 & 158.7  \\
\botrule
\end{tabular}
\label{tab:pairdist}
\end{table*}

\begin{table*}
\caption{Time matrix for baseball teams. The number at the $i$th row and $j$th column
denotes the travel time $t_{ij}$ from $i$ to $j$ in units of min, and the average traveling times ${\bar t_{i \bullet}}$
and ${\bar t_{\bullet j }}$ are defined by ${\bar t_{i\bullet}} \equiv (1/8)\sum_{j=1}^8 t_{ij}$ and
${\bar t_{\bullet j}} \equiv (1/8)\sum_{i=1}^8 t_{ij}$. }
\begin{tabular}{c|c|c|c|c|c|c|c|c}
\toprule
from\textbackslash to & 1 (SS)  &  2 (SK)  &  3 (LT)  &  4 (KA)  &  5,6 (DS,LG)  &  7 (HH)  &  8 (NX)   & ${\bar t_{i \bullet}}$ \\  
\colrule
 1 (SS)  &   0 & 239 & 105& 222 & 215 & 139& 242 & 172\\ \colrule                        
 2 (SK)  & 233 &   0 & 310& 234 & 77  & 161&  49 & 143\\ \colrule                        
 3 (LT)  & 113 & 325 &   0& 208 & 307 & 221& 336 & 227\\ \colrule                        
 4 (KA)  & 217 & 255 & 206&   0 & 239 & 149& 271 & 197\\ \colrule                        
 5,6 (DS,LG) & 211 &  74 & 288& 230&   0  & 148&  39 & 124\\ \colrule                        
 7 (HH)  & 129 & 180 & 209& 151 & 161 &   0& 189 & 148\\ \colrule                        
 8 (NX)  & 231 &  47 & 311& 229 & 38  & 162&   0 & 132\\
\colrule
${\bar t_{\bullet j}}$ &  168 &  149  & 215 & 188 & 130 & 141 & 146    \\
\botrule
\end{tabular}
\label{tab:pairtime}
\end{table*}

\begin{figure*}
\includegraphics[width=0.90\textwidth]{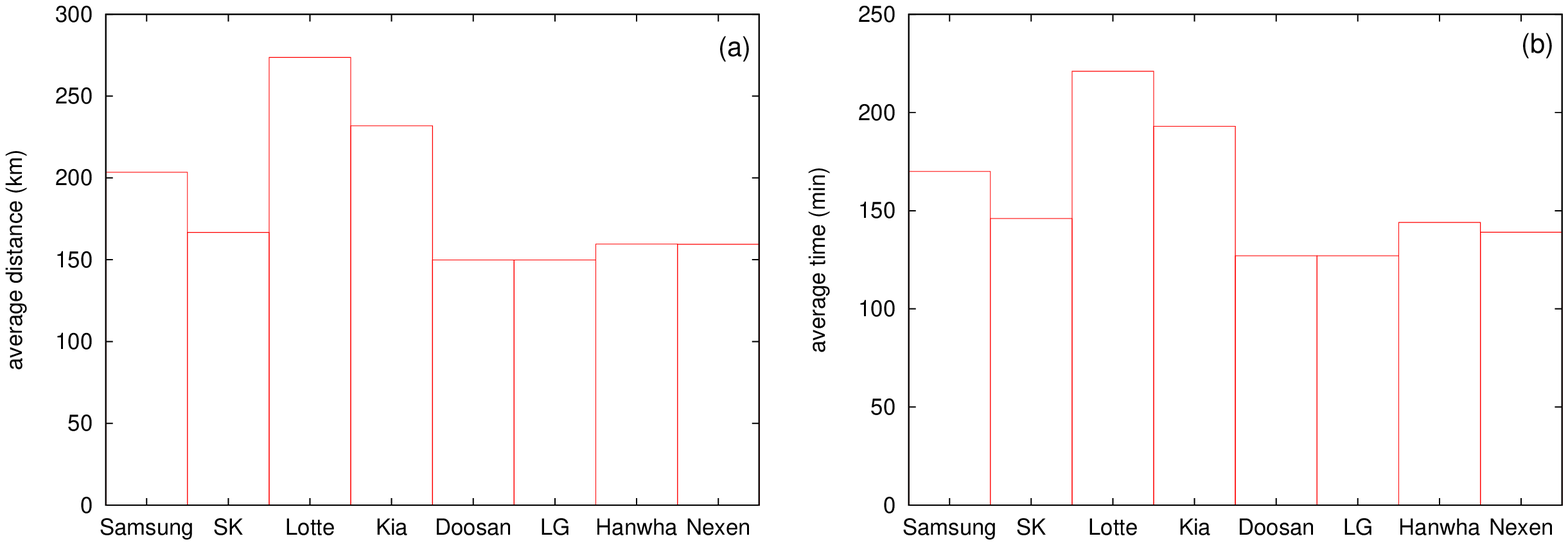}
\caption{(Color online) Average (a) distance ($\equiv (d_{i\bullet} + d_{\bullet i})/2$) and 
(b) time ($\equiv (t_{i\bullet} + t_{\bullet i})/2$) from Table~\ref{tab:pairdist} and~\ref{tab:pairtime}.
Clearly shown is that the team Lotte (see Fig.~\ref{fig:stadium}) has larger average distance and time
to travel than other teams. 
} 
\label{fig:avgdt}
\end{figure*}

\section{Scheduling of Korean Baseball League}
\label{sec:kborule}

In Korean Baseball League (KBL), there are now eight teams as shown in Table~\ref{tab:teams}.
Two teams (5 and 6, Doosan Bears and LG Twins)  based on Seoul, the capital
of South Korea with over 10 million population, share the same stadium Jamsil.
The locations of home stadiums are denoted in the form of (latitude, longitude)
in Table~\ref{tab:teams}. We also display the home positions of baseball teams 
on the map of South Korea in Fig.~\ref{fig:stadium}, which clearly shows that teams are not uniformly 
scattered across the country and that there are more teams around Seoul area.
Since baseball teams usually travel by bus (in Monday afternoon and Thursday night), 
we measure the distance
for a given pair of two teams by the output from the path finding service
provided by a company in Internet. More specifically, we type in the location
of the stadium of the team $i$ as the departure position, and then that of the team $j$
as the destination position. The Web service provided by the company Naver ({\tt http://www.naver.com})
then returns us the distance $d_{ij}$ and the estimated time $t_{ij}$ for the travel path
from $i$ to $j$, as shown in Tables~\ref{tab:pairdist} and~\ref{tab:pairtime}, 
respectively. It is to be noted that $d_{ij} \neq d_{ji}$, and $t_{ij} \neq t_{ji}$, 
although the differences are not so substantial. 
Figure~\ref{fig:avgdt} displays (a) $\bar{d_i} \equiv (1/16)\sum_{j=1}^8 ( d_{ij} + d_{ji} )$ and 
(b) $\bar{t_i} \equiv (1/16)\sum_{j=1}^8 ( t_{ij} + t_{ji} )$, respectively.
We note that Lotte, Kia, and Samsung have larger average distances and times than 
other teams. The shorter average distances for Doosan, LG, SK, and Nexen are because
of their locations near Seoul, and Hanwha has a short average distance because it
is located in the middle of the country. From Fig.~\ref{fig:avgdt}, we recognize that
if the tournament scheduling is made without any consideration of the inequality
of the average distances, some teams must travel longer distances spending longer
times in traffic than other teams, and thus those teams might have less chance to
win the games. The original formulation of TTP was made to minimize the total travel
distances of all teams. However, in view of the unequal distance distribution in KBL
(see Fig.~\ref{fig:avgdt}), it is natural to shift the optimization focus toward
the fairness of the tournament schedule, which will be the main research theme
of the present work. 

\begingroup
\squeezetable
\begin{table*}
\caption{Baseball schedule for 2012 season by KBO. Fixture of all 500 games is
given. For acronyms of team names and stadiums, see Table~\ref{tab:teams}. 
Each pair of two teams in the form XX/YY 
in the table denotes three consecutive games, one per day, with XX (YY) the home (away) team.
See text for the rules set by KBO. To compensate the longer travel distance of
Lotte, KBO allows the team to have 9 consecutive away games, as underlined in
the schedule (in the 11th and the 12th weeks).} 
\begin{tabular}{|c||c|c|c|c|c|c|c||c|c|c|c|c|c|c|c|c}
\toprule
 & \multicolumn{7}{c||}{Tuesday - Thursday} & \multicolumn{7}{c|}{Friday - Sunday} \\
\colrule
Week & Jamsil & Mokdong & Munhak & Daegu & Gwangju & Daejeon & Sajik  & Jamsil & Mokdong & Munhak & Daegu & Gwangju & Daejeon & Sajik \\
 \colrule
1 & \multicolumn{7}{c||}{No game}& 
{\bf DS/NX}\footnotemark[1]\footnotetext[1]{The schedule of the opening week is fixed.}	& -	& {\bf SK/KA}\footnotemark[1]& {\bf SS/LG}\footnotemark[1]& -	& -	& {\bf LT/HH}\footnotemark[1]\\
2&LG/LT	& NX/SK	& -	& -	& KA/SS	& HH/DS	& - &
LG/KA	& -	& SK/HH	& SS/NX	& -	& -	& LT/DS\\
3&DS/SS	& NX/KA	& -	& -	& -	& HH/LG	& LT/SK &
LG/SK	& NX/DS	& -	& -	& KA/LT	& HH/SS	& -\\
4&LG/NX	& -	& SK/DS	& SS/LT	& KA/HH	& -	& - &
DS/KA	& -	& SK/SS	& -	& -	& HH/NX	& LT/LG\\
5&LG/HH	& NX/LT	& -	& SS/DS	& KA/SK	& -	& - &
{\bf LG/DS}\footnotemark[2]\footnotetext[2]{Around the Children's Day, home stadiums are fixed.}		& -	& {\bf SK/LT}\footnotemark[2]	& {\bf SS/HH}\footnotemark[2]	& {\bf KA/NX}\footnotemark[2]	& -	& -\\
6&DS/SK	& NX/LG	& -	& -	& -	& HH/KA	& LT/SS & 
LG/SS	& -	& SK/NX	& -	& KA/DS	& HH/LT	& -\\
7&DS/HH	& -	& SK/LG	& SS/KA	& -	& -	& LT/NX &
DS/LG	& NX/SS	& -	& -	& -	& HH/SK	& LT/KA\\
8&LG/NX	& -	& SK/DS	& SS/LT	& KA/HH	& -	& - &
DS/LT	& NX/HH	& -	& SS/SK	& KA/LG	& -	& -\\
9&DS/KA	& NX/SK	& -	& -	& -	& HH/SS	& LT/LG &
LG/HH	& -	& SK/KA	& SS/DS	& -	& -	& LT/NX\\
10&DS/SK	& NX/LG	& -	& -	& KA/SS	& HH/LT	& - & 
LG/DS	& -	& SK/SS	& -	& -	& HH/NX	& LT/KA\\
11&LG/SK	& -	& -	& SS/HH	& KA/NX	& -	& LT/DS &
DS/SS	&  \underline{NX/LT}\footnotemark[3]\footnotetext[3]{Nine consecutive away games allowed for Lotte.}		& SK/HH	& -	& KA/LG	& -	& -\\
12&DS/NX	& -	&  \underline{SK/LT}\footnotemark[3]\	& SS/KA	& -	& HH/LG	& - &
 \underline{LG/LT}\footnotemark[3]\	& NX/SS	& -	& -	& KA/SK	& HH/DS	& -\\
13&LG/KA	& NX/DS	& -	& SS/SK	& -	& -	& LT/HH &
DS/LT	& -	& SK/LG	& SS/NX	& -	& HH/KA	& -\\
14&LG/SS	& NX/HH	& -	& -	& KA/DS	& -	& LT/SK & 
LG/DS	& NX/KA	& -	& -	& -	& HH/SK	& LT/SS\\
15 &DS/HH	& -	& SK/NX	& SS/LG	& KA/LT	& -	& - &
LG/NX	& -	& SK/DS	& SS/KA	& -	& -	& LT/HH \\
16&LG/SK	& NX/LT	& -	& -	& KA/DS	& HH/SS	& - &
\multicolumn{7}{c|}{All Star Game}\\
17&DS/LG	& NX/KA	& -	& SS/SK	& -	& HH/LT	& - &
DS/LT	& NX/SS	& SK/LG	& -	& KA/HH	& -	& -\\
18&LG/HH	& -	& SK/NX	& SS/DS	& -	& -	& LT/KA &
DS/KA	& NX/LG	& -	& -	& -	& HH/SK	& LT/SS\\
19&LG/LT	& -	& SK/SS	& -	& KA/NX	& HH/DS	& - &
DS/SK	& NX/HH	& -	& SS/LG	& KA/LT	& -	& -\\
20&LG/KA	& NX/DS	& -	& SS/HH	& -	& -	& LT/SK &
DS/SS	& -	& SK/KA	& -	& -	& HH/LG	& LT/NX\\
21&DS/NX	& -	& SK/HH	& SS/LT	& KA/LG	& -	& - &
LG/SS	& NX/SK	& -	& -	& -	& HH/KA	& LT/DS\\
22&DS/LG	& -	& SK/LT	& -	& KA/SS	& HH/NX	& - &
DS/HH	& -	& -	& SS/NX	& KA/SK	& -	& LT/LG\\
\botrule
\end{tabular}
\label{tab:KBOschedule}
\end{table*}
\endgroup

We next describe the tournament scheduling scheme of KBO in detail. At the time
of writing of the present paper (year 2012), each team in one season plays 133 games, which corresponds
to 19 $(=133/7)$ games for each pair of two teams, making the total number of games 
532 $(=133 \cdot  8/2)$. The scheduling constraints set by KBO are as follows:
\begin{enumerate}

\item The number of home/away games are either 66/67 or 67/66 (with $66+67 = 133$), 
respectively. If a team had 66 (67) home games in the last season, 67 (66) 
home games are assigned. In 2011, Kia, LG, Hanwha, and Nexen had 67 home games
and Samsung, SK, Lotte, and Doosan had 66 home games. In 2012, therefore, the four
former teams should have 66 home games and the latter 67 home games.

\item There is no game on Monday. From Tuesday to Thursday two teams play three
games in a row at the same stadium (home stadium of one of the two teams), and
then move to other stadium to play three more games in a row from Friday to
Sunday. 

\item The first week of the season is an exception: New season starts with an opening ceremony on Saturday. Each team
plays two games on Saturday and Sunday without changing counterpart.
The four stadiums of the first week are the homes of the top four teams 
of the season 2010 (SK, Samsung, Doosan, and Lotte). The final 
rank in 2010 determines the counterpart of each game 
[rank 1 (SK) plays with rank 5 (Kia), 2 (Samsung) with 6 (LG), 3 (Doosan) with 7 (Nexen), 
and 4 (Lotte) with 8 (Hanwha)].

\item Among all 532 games, the game fixture for the first 500 games
are fixed before the season starts (see Table~\ref{tab:KBOschedule}). 
First week has 8 games, and from
the second week 24 games are held every week. The third week
of July (the 16th week of the season) is another exception and regular season matches are held only
for the first half of the week. On Saturday (July 21) there
will be a special event so-called "All Star Game".
Accordingly, after 22 weeks, all 500 [$= 8$ (the 1st week) + $12$ (the 3rd week of July)
+$24\cdot 20$ (regular weeks)] games are played, and remaining 32 games start from September 2.
Fixture of those remaining 32 games will be determined during the season, 
reflecting the unavoidable changes made by e.g., canceled game due to heavy rain.

\item Around the Children's' Day in Korea (May 5), the most popular day for kids
to watch a baseball game in a stadium with family, four stadiums for three consecutive
games are chosen as the homes of Samsung, SK, Kia, and LG. Every second year,
the four teams alternate so that Doosan, Nexen, Lotte, and Hanwha will be the ones
in the 2013 season. 

\item  No team is allowed to play more than 6 home games consecutively.

\item No team is allowed to play more than 6 away games consecutively,
with one exception:
KBO noticed the longer travel distance of Lotte, and allows
the team to play 9 away games in a row, before the 500th game of
the season (from June 15 to 24 in the 2012 season). In Table~\ref{tab:KBOschedule},
we underline those 9 games in the 11th and 12th weeks of the schedule.
\end{enumerate}

\begin{table}
\caption{Total traveling distance and time for the 2012 schedule in 
Table~\ref{tab:KBOschedule}.}
\begin{tabular}{c|c|c|c}
\toprule
\# & team  & distance (km) & time (min) \\
\colrule
1 &Samsung   & 9086.9 & 7484 \\
2 &SK        & 6714.7 & 5881 \\
3 &Lotte     & 9204.9 & 7594 \\
4 &Kia       & 8311.1 & 7007 \\
5 &Doosan    & 6795.0 & 5692 \\
6 &LG        & 5538.0 & 4764 \\ 
7 &Hanwha    & 7017.0 & 6154 \\ 
8 &Nexen     & 6552.4 & 5645 \\
\colrule
& average & 7402.5 & 6278  \\
\botrule
\end{tabular}
\label{tab:KBOtotal}
\end{table}

The above scheduling rules set by KBO show that the organization is trying
to provide equal bases for all the teams. Especially, the rule~7 makes
Lotte an exception to compensate the team's longer distance as 
shown in Tables~\ref{tab:pairdist} and~\ref{tab:pairtime}, and Fig.~\ref{fig:avgdt}.
We use the KBO
schedule for the 2012 season in Table~\ref{tab:KBOschedule}, to compute the total
traveling distance and time of the season. For example, the team SK plays
three home games at Munhak stadium in the 1st week ($\sum d=0$), 
and travels to Mokdong stadium to play three away games with NX ($\sum d= d_{28}$),
and then back to Munhak stadium for three home games with HH ($\sum d=d_{28} + d_{82}$)
in the 2nd week. SK keeps traveling following the schedule in Table~\ref{tab:KBOschedule} 
so that its total traveling distance is $\sum d=d_{28} + d_{82} + d_{23} +
d_{35} + d_{52} + \cdots + d_{24}$. When there are consecutive away games before and 
after Monday, we assume that the team travels directly from one place to other,
without having a break on Monday at the team's home.
We also compute the total traveling time similarly to 
the total traveling distance, and list the both in Table~\ref{tab:KBOtotal}. 
Clearly seen in Table~\ref{tab:KBOtotal} is that the above
KBO rules do not successfully reduce the unequal distribution of traveling distances
and times. For example, even though Lotte is given an exception (the rule 7)
the team travels much longer distance (66\%) than LG. Another interesting
observation is that although Doosan and LG are located at the same home position,
the travel distances are quite different: Doosan travels 23\% longer 
distance compared to LG. 

\begin{figure*}
\includegraphics[width=0.98\textwidth]{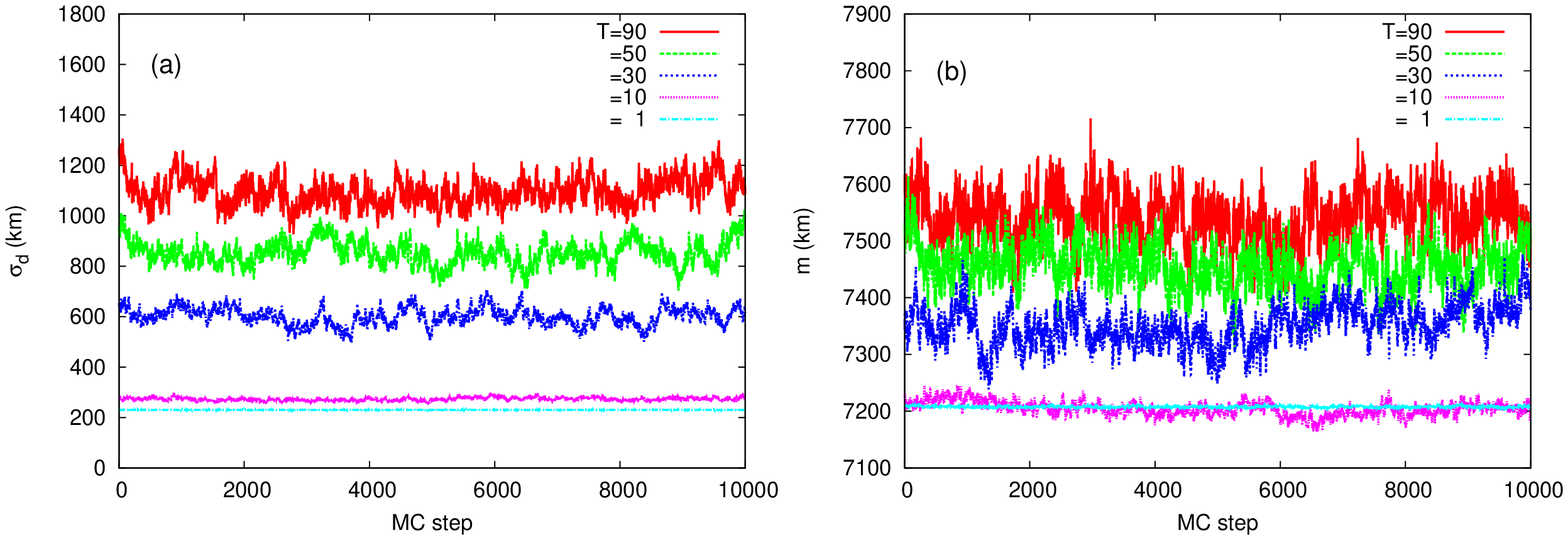}
\caption{(Color online) Evolution of (a) the standard deviation of 
traveling distance $\sigma_d$ and (b) the average traveling distance $m$
versus Monte-Carlo time step. We start from a high enough temperature $T=100$
and decrease $T$ slowly after 10,000 MC updates at each temperature.
As $T$ is lowered from $T=90$, to 50, 30, 10, and 1 (from top to bottom curves),
$\sigma_d$ systematically decreases. (b) Although not explicitly intended
$m$ is also found to decrease with $T$ for $T \gtrsim 10$.} 
\label{fig:MCseries}
\end{figure*}

\begingroup
\squeezetable
\begin{table*}
\caption{Optimized schedule in this work with maximum fairness in traveling distances.} 
\begin{tabular}{|c||c|c|c|c|c|c|c||c|c|c|c|c|c|c|c|c}
\toprule
 & \multicolumn{7}{c||}{Tuesday - Thursday} & \multicolumn{7}{c|}{Friday - Sunday} \\
 \colrule
 Week & Jamsil & Mokdong & Munhak & Daegu & Gwangju & Daejeon & Sajik  & Jamsil & Mokdong & Munhak & Daegu & Gwangju & Daejeon & Sajik \\
  \colrule
1 & \multicolumn{7}{c||}{No game} & {\bf DS/NX}\footnotemark[1]\footnotetext[1]{The schedule of the opening week is fixed.} & -  & {\bf SK/KA}\footnotemark[1] & {\bf SS/LG}\footnotemark[1] & -  & -  & {\bf LT/HH}\footnotemark[1] \\
2 & DS/KA & NX/LG & SK/HH & SS/LT & -  & -  & -  & LG/SS & -  & -  & -  & KA/DS & HH/NX & LT/SK \\
3 & DS/SK & NX/LG & -  & -  & KA/SS & -  & LT/HH & LG/NX & -  & SK/DS & SS/KA & -  & HH/LT & -  \\
4 & DS/LG & -  & -  & SS/NX & KA/LT & HH/SK & -  & DS/KA & NX/HH & SK/SS & -  & -  & -  & LT/LG \\
5 & LG/SK & NX/KA & -  & -  & -  & HH/SS & LT/DS & {\bf LG/DS}\footnotemark[2]\footnotetext[2]{Around the Children's Day, home stadiums are fixed.} & -  & {\bf SK/LT}\footnotemark[2] & {\bf SS/HH}\footnotemark[2] & {\bf KA/NX}\footnotemark[2] & -  & -  \\
6 & DS/LT & -  & -  & SS/SK & KA/LG & HH/NX & -  & LG/KA & NX SS & SK HH & -  & -  & -  & LT/DS \\
7 & DS/LG & NX/KA & SK/SS & -  & -  & -  & LT/HH & DS/SK & -  & -  & SS/LG & KA/LT & HH/NX & -  \\
8 & LG/KA & NX/DS & SK/SS & -  & -  & HH/LT & -  & DS/HH & -  & SK/KA & SS/NX & -  & -  & LT/LG \\
9 & DS/SK & NX/LG & -  & -  & KA/HH & -  & LT/SS & LG/NX & -  & SK/DS & SS/LT & -  & HH/KA & -  \\
10 & LG/HH & NX/SK & -  & SS/DS & -  & -  & LT/KA & DS/SS & -  & -  & -  & KA SK & HH/LG & LT/NX \\
11 & DS/LT & NX/SS & SK/LG & -  & -  & HH/KA & -  & LG/SK & NX/LT & -  & SS/HH & KA/DS & -  & -  \\
12 & DS/KA & -  & SK/NX & -  & -  & HH/SS & LT/LG & DS/LG & NX/HH & -  & SS/KA & -  & -  & LT/SK \\
13 & LG/LT & -  & -  & SS/SK & KA/NX & HH/DS & -  & DS/NX & -  & SK/LT & -  & KA/SS & HH/LG & -  \\
14 & LG/KA & NX/DS & -  & SS/HH & -  & -  & LT/SK & LG/SS & -  & SK/NX & -  & KA/DS & HH/LT & -  \\
15 & DS/HH & NX/SK & -  & SS/LG & KA/LT & -  & -  & LG/NX & -  & SK/DS & -  & -  & HH/KA & LT/SS \\
16 & DS/LT & -  & SK/LG & SS/NX & KA/HH & -  & -  & \multicolumn{7}{c|}{All Star Game}                    \\
17 & LG/HH & NX/LT & -  & SS/DS & KA/SK & -  & -  & LG/SK & NX/DS & -  & -  & -  & HH/SS & LT/KA \\
18 & DS/HH & -  & SK/LG & SS/KA & -  & -  & LT/NX & LG/LT & NX/SS & -  & -  & KA/SK & HH/DS & -  \\
19 & DS/SS & -  & SK/LT & -  & KA/NX & HH/LG & -  & LG/DS & NX/HH & -  & SS/SK & -  & -  & LT/KA \\
20 & LG/SS & -  & SK/NX & -  & KA/HH & -  & LT/DS & DS/NX & -  & -  & SS/LT & KA/LG & HH/SK & -  \\
21 & LG/DS & NX/KA & SK/HH & -  & -  & -  & LT/SS & LG/HH & -  & SK/KA & SS/DS & -  & -  & LT/NX \\
22 & DS/SS & NX/LT & -  & -  & KA/LG & HH/SK & -  & LG/LT & NX/SK & -  & -  & KA/SS & HH/DS & -  \\
\botrule
\end{tabular}
\label{tab:Ourschedule}
\end{table*}
\endgroup

\section{Formulation}
\label{sec:form}
The total traveling distance of the team $i$ is given by
\begin{equation}
d_i  = \sum_{(k,l)} d_{kl} ,
\end{equation}
where the sum is over the traveling path for a given schedule, and
$d_{kl}$ is the distance from $k$ to $l$ in Table~\ref{tab:pairdist}.
Differently from the usual
TTP in which the objective function to minimize is $\sum_i d_i$, we in this
work propose to minimize the unfairness of the schedule, measured by
the standard deviation of $d_i$. In analogy to the statistical physics
problem finding the ground state of a system, we call it as ``energy''
$E$ of the system, although the term does not mean much in 
the context of baseball game scheduling. Our objective function to minimize
in the present work is given by
\begin{equation}
E = \sqrt{\frac{1}{8} \sum_{i=1}^8 (d_i - m)^2} = 
\sqrt{\langle d_i^2 \rangle - \langle d_i \rangle^2} \equiv \sigma_d,
\label{eq:E}
\end{equation}
where the mean $m \equiv \langle d_i \rangle \equiv (1/8)\sum_{i=1}^8 d_i$, and 
$\langle d_i^2 \rangle \equiv (1/8)\sum_{i=1}^8 d_i^2$.
In contrast to the original TTP to minimize $m$, our purpose here is to minimize
$E$. In the ideal situation of $E=0$, all teams travel equal total distances, 
while the larger is $E$, the more unfair the schedule becomes.
We also use the standard deviation $\sigma_t$ of the travel time $t_i \equiv \sum_{(k,l)} t_{kl}$
with $t_{kl}$ being the travel time from $k$ to $l$ in Table~\ref{tab:pairtime}, 
to define the energy function $E$.

In statistical physics, finding of the ground state configuration of a model
system is the one of the most well-developed topics. The standard way 
to tackle the issue is to set up an energy function of the given model system
and minimize the energy in a systematic way. In this regard, the most standard
methodology is to use the Monte-Carlo (MC) simulation technique~\cite{MCbook}.
The standard Metropolis algorithm for local update is adopted: The local MC try
with the energy difference $\Delta E$ is accepted if $\Delta E \leq 0$.
Otherwise (i.e., if $\Delta E > 0$),  the MC try is accepted at the probability
of $e^{-\Delta E/T}$ with the temperature $T$.
The Metropolis MC method is then combined with the simulated annealing technique. 
In more detail, we start from a high temperature, 
and decrease the temperature slowly until the zero temperature is reached. 
The similar MC method with the simulated annealing technique has been 
applied for the TSP~\cite{kirkpatrick}. In Ref.~\onlinecite{leechoi},
a microcanonical MC method combined with an annealing technique 
called the entropic annealing method has also been applied for the TSP.
In the present study, we do not aim at developing a more advanced methodology
in finding the fairest tournament schedule, but we simply apply the well-known
simulated annealing MC method to tackle the problem.  
Accordingly, we do not claim that we have found the fairest schedule of the KBL,
but rather like to emphasize that a simple application of the
well-known methodology in statistical physics 
can significantly increase the fairness of the schedule of the KBL.

The other important issue to discuss is what are the local update rules allowed
in the MC simulation. In Ref.~\onlinecite{TTP:SA}, all five different ways of
update are proposed: SwapHomes, SwapRounds, SwapTeams, PartialSwapRounds, and
PartialSwapTeams. It has been shown that the first three swapping methods are
not sufficiently efficient to explore the entire configuration
space~\cite{TTP:SA,TTP:ACO}.  In this work, we only use SwapRounds and
SwapHomes for updates due to their simplicity: In the former we pick two different rounds, e.g.,
Fri-Sun of the week 7 and Tue-Thu of the week 6, and swap the two entirely. In
the SwapHomes update, we pick two pairs, e.g., NX/SK and SK/NX at two different
rounds Tue-Thu of the week 2 and Fri-Sun of the week 6, and then swap the two.
These two update methods (SwapRounds and SwapHomes) are the simplest ones among
the five in Ref.~\onlinecite{TTP:SA} and it is clear that the two update
methods preserve the KBO rules explained in Sec.\ref{sec:kborule} if
the schedule in Table~\ref{tab:KBOschedule} is used as an initial
configuration of MC simulations.

\section{Results}
\label{sec:res}

We start from a sufficiently high temperature $T=100$ with the KBO schedule
in Table~\ref{tab:KBOschedule} as an initial condition 
and decrease $T$ slowly with the temperature step $\Delta T = 1.0$ until $T=0$
is arrived. For 22 weeks of the season in Table~\ref{tab:KBOschedule}, 
there are all 42 rounds, with two rounds per week except for the 1st and 16th weeks.
At each temperature, we perform 10,000 MC steps, with one MC step
composed of one SwapRounds update and one SwapHomes updates per round. 
Our update rules do not change the number of home games and away games for each team, and thus
the rule~1 of KBO in Sec.~\ref{sec:kborule} is automatically obeyed.
The schedule of the first week is fixed (rule 3) in our simulation,
and the ``All Star Game'' in the 16th week is also fixed (see rule 4). 
No teams are allowed to have more than six consecutive home games (rule 6), 
nor six consecutive away games (rule 7).
The exception made for Lotte (rule 7 in Sec.~\ref{sec:kborule}) is not
necessary in our simulation. The rule 5 on
the home stadiums around the Children's Day is also obeyed during simulations.
Consequently, all rules (except the unnecessary exception for Lotte in rule 7) are equally
obeyed for all teams during our simulations at any temperature, and the ground state
as an outcome of the simulated annealing MC simulations gives us the
tournament schedule that is expected to be as fair as possible in terms
of the travel distance (for $E = \sigma_d$)  or the travel time ($E = \sigma_t$).
We repeat the same annealing procedure 10 times  for the fairness of traveling distances 
($E = \sigma_d$) and once for the fairness of traveling times ($E = \sigma_t$).
The results are almost the same for $E = \sigma_d$ and $E = \sigma_t$, and we below
mostly present our results for the former.

In Fig.~\ref{fig:MCseries}, we show the evolution of (a) $\sigma_d$ and
(b) $m$ in Eq.~(\ref{eq:E}) during MC simulated annealing. As the temperature
is lowered, it is shown that our simulation successfully lowers
the energy, or increase fairness in the travel distance. Although
our simulation tries to minimize $\sigma_d$, 
$m$ is also found to decrease as a byproduct of annealing as shown in 
Fig.~\ref{fig:MCseries}(b). Ten independent simulated annealing with 
$E = \sigma_d$ give us ten ground state configurations, and the one
with the lowest energy has the configuration tabulated in 
Table~\ref{tab:Ourschedule}. We emphasize that our optimized 
baseball schedule keeps all the scheduling constraints by KBO,
but with significantly enhanced fairness in travel distances.

\begin{table}
\caption{Total traveling distance of each team is listed for the optimized 
schedule in Table~\ref{tab:Ourschedule}, obtained 
from MC simulated annealing with the fairness of traveling distances
as the objective function. The total traveling times obtained when
the standard deviation of the traveling times is chosen as the 
energy function are also listed.
Note that the average distance and time are smaller
than the ones in actual KBO schedule in Table~\ref{tab:KBOtotal}. 
See text for more details. }
\begin{tabular}{c|c|c|c}
\toprule
\# & team  & distance (km) & time (min) \\
\colrule
1 &Samsung   & 7185.6 	&	6103  \\
2 &SK        & 7163.5 	&	6071  \\
3 &Lotte     & 7252.7 	&	6233  \\
4 &Kia       & 7205.8 	&	6139  \\
5 &Doosan    & 6980.0 	&	5986  \\
6 &LG        & 6866.3 	&	5852  \\ 
7 &Hanwha    & 6898.5 	&	6064  \\ 
8 &Nexen     & 7183.4 	&	6074  \\
\colrule
& average   & 7092.0 & 6065  \\
\botrule
\end{tabular}
\label{tab:Ourtotal}
\end{table}

\begin{figure*}
\includegraphics[width=0.98\textwidth]{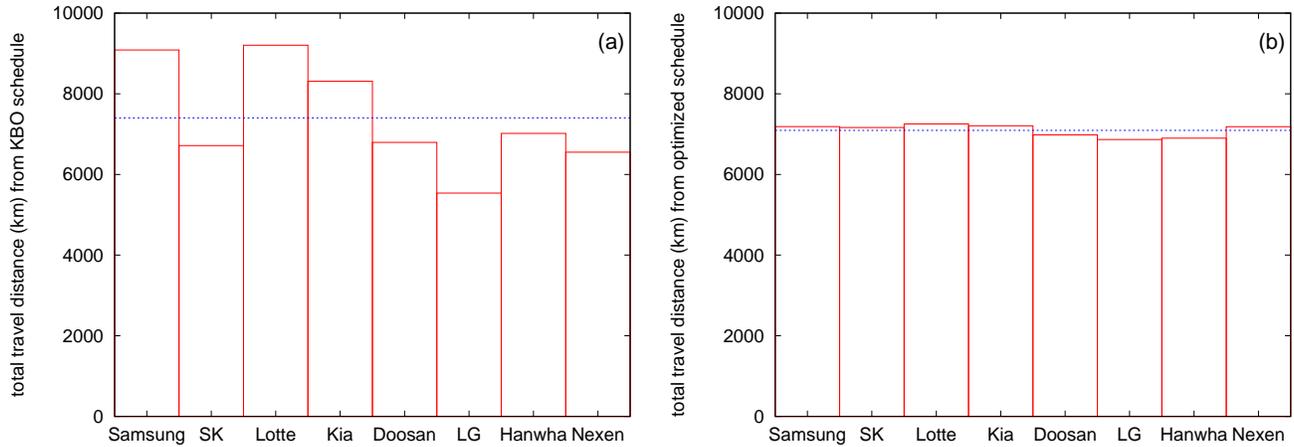}
\caption{(Color online) (a) Total traveling distance 
for the KBO 2012 schedule in Table~\ref{tab:KBOschedule}.
Total traveling distance is quite different for each team, with
Lotte the worst and LG the best. 
(b) Total traveling distance from the optimized schedule 
in Table~\ref{tab:Ourschedule}.  Our optimized schedule provides much better fairness in view of
the distribution of traveling distances.  
The horizontal dotted lines in (a) and (b) are average total
travel distance over eight teams. 
Interestingly, the average distance is shorter for the optimized
schedule than the one in the actual KBO schedule, which means our suggested schedule
is better than the KBO schedule in terms of both fairness and effectiveness.} 
\label{fig:traveldist}
\end{figure*}

In Table~\ref{tab:Ourtotal}, we list the total traveling distances
of teams for the ground state configuration in Table~\ref{tab:Ourschedule}.
In comparison to the corresponding values for the actual schedule
run by KBO in Table~\ref{tab:KBOtotal}, our optimized schedule clearly
has more even distribution of travel distances. 
In Table~\ref{tab:Ourtotal}, we also include the result from 
our MC simulated annealing with $E = \sigma_t$, i.e., 
the fairness in traveling times as the objective function.
The traveling time of each team has much smaller variance
in comparison to the original schedule by KBO.

To summarize our main findings, we plot in Fig.~\ref{fig:traveldist}
the traveling distances for 
(a) the KBO schedule in Tables~\ref{tab:KBOschedule} and~\ref{tab:KBOtotal}
and (b) our optimized schedule in Tables~\ref{tab:Ourschedule} and~\ref{tab:Ourtotal}.
Clearly revealed is that our optimization scheme results in much better
fairness in travel distances. 
The corresponding plot for the
total traveling time looks very similar to Fig.~\ref{fig:traveldist} 
and thus not included in the present paper. 
Although the team Lotte still has the 
longest distance to travel, the difference to LG is now only about 6\%
(compared to 66\% in original KBO schedule).
Also, the difference between  LG and Doosan shrinks to 2\%
(compared to 23\% in KBO schedule).
We also report somehow unexpected result of our scheme: Although
not intended, the total sum of all the traveling distances of 
eight teams in our optimized schedule has {\it smaller} value
than the one for the original KBO schedule. Accordingly,
we conclude that we manage to find better schedule 
in terms of both fairness and effectiveness, in comparison
to the KBO schedule.

\section{Summary and Discussion}
\label{sec:sum}
We have studied numerically the traveling tournament problem in Korean Baseball League.
Instead of using the total sum of traveling distances as the objective function to 
minimize, we focus on the fairness in the traveling distances of baseball teams.
More specifically, we use the standard deviation of the traveling distance 
as the energy function in statistical mechanics, and apply the simulated
annealing Monte-Carlo method to find the ground state configuration.
The resulting timetable for the baseball games for year 2012 is shown to be better
than the schedule posted by the Korean Baseball Organization, in terms of
both the fairness, and the effectiveness. In other words, our optimized schedule
yields more equal traveling distances across teams, and the total sum of
total traveling distances is shorter than the corresponding value of the posted
schedule by the organization.

We admit that there could be some other implicitly posed constraints
not reflected in our simulations. However, if those constraints are 
explicitly revealed to public, our methodology can be applied with
little changes in the MC program. We strongly believe
that statistical physics approach like presented in this work has a
broad applicability in a broad range of problems we encounter
everyday, and hope to get encountered by more interesting problems in the
real world of society.

\begin{acknowledgments}
This work was done as a team project in the 9th KIAS-APCTP Winter School on
Statistical Physics in Korea (2012).  B.J.K. was supported by the National Research Foundation of 
Korea (NRF) grant funded by the Korea government (MEST) (No. 2011-0015731).
\end{acknowledgments}

%\bibliographystyle{apsrev4-1}
%\bibliography{baseball}

%merlin.mbs apsrev4-1.bst 2010-07-25 4.21a (PWD, AO, DPC) hacked
%Control: key (0)
%Control: author (72) initials jnrlst
%Control: editor formatted (1) identically to author
%Control: production of article title (-1) disabled
%Control: page (0) single
%Control: year (1) truncated
%Control: production of eprint (0) enabled
%

\end{document}